\documentstyle [aps,prl,psfig,multicol,epsf]{revtex}
\begin{document}
\title{High-resolution Compton Scattering as a Probe of \\
Fermiology and Electron Correlation Effects}
\author{A. Bansil and B. Barbiellini}
\address{
Physics Department,
Northeastern University,
Boston, MA 02115, 
USA}
\maketitle
\begin{abstract}
\centerline{{\bf ABSTRACT\\}}
Compton scattering is one of the few spectroscopies which 
directly probes the ground state momentum density in materials. 
Recent progress in synchrotron light sources 
has brought a renewed interest in the technique 
as a tool for investigating fermiology related issues on the 
one hand, and as a unique window on hitherto inaccessible 
correlation effects in the electron gas on the other.
We provide an overview of some of our recent work on a variety of metals and 
disordered alloys, including Li, $\rm Li_{1-x}Mg_x$, Be, Cu, 
$\rm Ni_{75}Cu_{25}$, 
$\rm Ni_{75}Co_{25}$, Al and Al-Li.
\end{abstract}
\vskip 1.5cm

\begin{multicols}{2}
\section{INTRODUCTION}
Compton scattering is the inelastic scattering of x-rays in the deep 
inelastic regime (i.e. when the energy and momentum transfered in 
the scattering process are large).\cite{williams77,cooper85,mijnarends95}
It is one of the few spectroscopies which is capable of directly probing 
the electronic ground state in materials. The measured double-differential 
cross-section describing the energy and angular distribution of 
scattered radiation-- usually referred to as the Compton profile (CP), can 
be interpreted as
\begin{equation} 
J(p_z)=\int\int n({\bf p})dp_x dp_y~,
    \label{Eq1}                
\end{equation}
where $n({\bf p})$ is the ground-state electron momentum density. 
The theoretical analysis of the Compton spectra in the literature 
is largely based on the expression of $n({\bf p})$  within the 
independent particle model:
\begin{equation}  
    n({\bf p}) = \frac{1}{(2\pi)^3}      
    \sum_{{\bf k} \nu}\int
    |\psi_{{\bf k} \nu}({\bf r})\exp(i{\bf p \cdot r})d{\bf r}|^2,  
    \label{Eq2}                
\end{equation}
where $\psi_{k \nu}({\bf r})$ denotes an electron wavefunction 
in the state ${\bf k}$ and
band $\nu$, and the summation is over all occupied states. 
When the right 
side of Eq. (\ref{Eq2}) is evaluated using the selfconsistent band theory 
based electron wavefunctions in the local density approximation
(LDA), and the sum is carried out to properly include the Fermi 
surface (FS) breaks in $n({\bf p})$, one obtains a fairly sophisticated 
description of the Compton spectrum. Eq. (\ref{Eq2})
makes it obvious that $n({\bf p})$ and thus the CP contains 
signatures of both, the shape of the occupied region of the k-space 
in the first and higher Brillouin zones (BZs), as well as the 
nature of the electron wavefunctions. 
Since $n({\bf p})$ in Eq. (\ref{Eq1}) represents the full many-body ground 
state momentum density, a comparison of the measured CP
with that obtained via the LDA-band theory framework provides a 
direct measure of the extent to which the LDA one-particle wavefunctions
differ from the true quasiparticle states in the interacting electron
gas in the solid. 

The Compton technique offers a number of intrinsic advantages
over other k-resolved spectroscopies. 
Being essentially a ground state measurement, the Compton experiment
does not require long electron mean free paths as is the case in 
transport-type experiments such as the dHvA, the latter being limited
for this reason to systems with low defect and impurity concentrations.
Since no charged particles go in or out of the sample, Compton is
not complicated by surface effects present in photoemission or
electron scattering experiments, and is thus genuinely a bulk
probe. Moreover, because light couples weakly to the
electronic system, the disturbance of the electronic states one
is trying to measure is relatively smaller in Compton compared to positron
annihilation or photoemission-type spectroscopies which involve charged
particles. 

Historically, the capabilities of the Compton technique 
have been difficult to realize 
in practice because only a limited momentum resolution ($\approx$ 0.4 au) 
was possible with $\gamma$-ray sources. However, with the recent 
availability of second and third
generation high-energy, high-intensity synchrotron light sources, and 
improvements in detector technology it has become possible now to achieve 
momentum resolution of the order of 0.1 au in wide classes of materials, and 
of order 0.01 au in low-Z systems. These advances have 
sparked a renewed interest in the Compton technique 
as a tool for investigating fermiology related issues on the 
one hand, and as a unique window on hitherto inaccessible 
correlation effects in the electron gas on the other.\cite{bansil93,bankap97}

\section{AN OVERVIEW OF SELECTED RESULTS}

Within space limitations, we give below an overview of some of our recent 
high-resolution Compton studies in metals and disordered alloys. The list 
of references is meant to be minimal, but should 
serve as an entry point into other relevant literature.

\begin{enumerate} 

\item   Li has been the subject of much attention. 
Ref. \onlinecite{sakurai95} may be viewed as marking the beginning of the 
modern high-resolution Compton studies of solids. 
Ref. \onlinecite{sakurai95} showed that Compton
can provide accurate FS dimensions, at least in favorable cases. Furthermore, 
comparisons between highly accurate first-principles LDA-band-theory 
based predictions with the corresponding experimental profiles in Li 
showed clearly that there are systematic discrepancies between theory 
and experiment which could be attributed to correlation effects on 
the electron momentum density. 

\item  Considerable effort has been made to deduce the size of the break 
$Z_F$ in the momentum density at the Fermi momentum in Li. $Z_F$ is 
one of the most fundamental parameters which enters in the conventional 
theory of metals, and it has been investigated using a variety of 
theoretical approaches for many decades in the homogeneous electron gas. 
While there is still some uncertainty in analyzing the data, 
Ref. \onlinecite{schulke96} has 
estimated a near zero value of $Z_F$ in Li. This is most surprising and 
very far from theoretical predictions which yield $Z_F$ values ranging from 
about 0.65 to 0.82 in Li. 

\item  Several groups have attempted to ``reconstruct" the 3D-momentum 
density 
in Li from experimental CP's along several 
directions.\cite{schulke96,jap1,dob1}
In this connection we have computed highly accurate CP's along various 
groups of measured directions in Li to serve as a means of testing the 
accuracy of different resconstruction schemes. 

\item $\rm Li_{100-x}Mg_x$ disordered alloys have been investigated 
over the composition range $0\le x\le 40$.\cite{stutz99}
The experimental valence CP's, 
their second derivatives, and the associated directional anisotropies 
were compared with the corresponding parameter free KKR-CPA-LDA 
computations. Discrepancies between measurements and calculations were 
partly traced back to an inadequate treatment of correlation effects 
within the LDA. Curiously, the theoretical profiles display a better 
accord with experiments with increasing Mg concentration, hinting that 
the behavior of momentum density in Li may be idiosyncratic and not 
representative of metals more generally. 

\item  Be has been another system which has been the subject of several 
independent studies. The first high-resolution CP's reported in 
Ref. \onlinecite{hamalainen96}
showed the essential absence of some theoretically predicted FS
signatures in the data. However, this appears to have been due to a lack 
of adequate statistics in the data of 
Ref. \onlinecite{hamalainen96}. Subsequent Compton measurements 
on Be in Ref. \onlinecite{itou97} 
indicated no such serious discrepancies between theory and 
experiment. Ref. \onlinecite{huotari00}
has very recently carried out an extensive new study of 
Be single crystals using three different incident photon energies. 
Subtle differences between theoretical and experimental CP's are pointed 
out and it is suggested that a better treatment of the correlation effects 
in the {\it inhomogeneous} electron gas is needed to develop a 
satisfactory description of the momentum density in Be. 

\item  Al has been investigated quite extensively.\cite{suortti00,japal}
In order to deduce the value of $Z_F$ in Al, 
the data have been analyzed in terms of a simple model which involves 
$Z_F$ as a parameter. A good fit with the Compton data is obtained for 
$Z_F = $ 0.7 to 0.8.\cite{suortti00} 
Al has also been investigated via Compton experiments 
where the kinematics of the ejected electron is measured in coincidence 
with the scattered photon; in principle, such a ($\gamma$, e$\gamma$) 
experiment determines the 3D momentum density directly. The concidence 
Compton data on Al indicate a $Z_F$ value of about 0.7.\cite{suortti00}
These results suggest
that, in sharp contrast to the case of Li, the standard picture of the 
interacting electron gas is substantially correct in Al. 

\item  In Cu, high-resolution Compton measurements together with the 
corresponding highly accurate LDA-based computations along the three 
principal symmetry directions have been reported recently in 
Ref. \onlinecite{sakurai99}. 
A reasonable overall picture of the ground state momentum density in Cu
is adduced, and the deviations from the band theory predictions are 
generally found to be consistent with the high-resolution studies of Li 
and Be. 

\item  In Ref. \onlinecite{bansil98}, 
we have presented first-principles ferromagnetic electronic 
structures and spin-resolved Compton profiles along the three high-symmetry
directions in Ni, $\rm Ni_{75}Cu_{25}$ and 
$\rm Ni_{75}Co_{25}$ disordered alloys. 
The computations are based on the use of the charge- and spin-selfconsistent
KKR-CPA-LDA approach and involve no free parameters-- the lattice constants
in all cases were obtained by minimizing the total energy. The majority-spin
spectrum of Ni is found to undergo relatively small changes upon alloying 
with Cu or Co, and indeed the associated CP's in Ni, $\rm Ni_{75}Cu_{25}$ and 
$\rm Ni_{75}Co_{25}$ are very similar. High-resolution magnetic CP experiments
are not feasible at the moment, but the theoretical predictions are found 
to be in reasonable overall accord with the available $\gamma$-ray 
measurements. 

\item  Finally, we note that very recent experiments on Al-Li alloys with 
3$\%$ Li seem to indicate that Li-impurities when placed in the Al-matrix 
may be inducing some very interesting changes in correlations in the 
electronic system in their vicinity in the alloy.\cite{mat00}
This point is under further investigation. 
\end{enumerate}

A few words concerning the theoretical methodology for obtaining highly 
accurate Compton profiles in metals and alloys are appropriate. Even 
within the conventional LDA-based band theory framework, highly accurate 
computations of Compton profiles necessary for analyzing FS signatures
are quite demanding. The reason is that the integrand $n({\bf p})$
occurring in Eq. (\ref{Eq1}) possesses a fairly long range with
myriad FS breaks which must be intergrated using a fine momentum
space mesh. Further, $J(p_{z})$ must be evaluated over
a fine $p_{z}$-mesh extending to quite high $p_{z}$ values
to properly normalize the theoretical profile. 
We have however developed the methodology to carry out such 
computations in ordered as well as disordered many atom per unit cell 
systems, including the possibility of magnetic ordering.
\cite{kaprzyk90,bansil91,bansil92,bansil99}
The relevant Green function formulation for treating momentum density and 
CP's in disordered alloys is given in Refs. \onlinecite{mijnarends76,bansil81}. 

In conclusion, it is clear that high-resolution Compton studies
can provide new insights into a variety of issues related to 
fermiology and correlation effects in metals and alloys. 
Of particular interest will be problem 
areas where the conventional k-resolved spectroscopies (dHvA, 
photoemission and positron annihilation) are difficult to apply due to 
surface or defect problems, as is the case for example 
in the oxides generally. Phase transformations and electronic structure 
and correlation effects under high pressure could be another niche area 
for Compton scattering since it is difficult to take charged particles 
in or out of a pressure cell.

\acknowledgements
It is a pleasure to thank Stanislaw Kaprzyk and 
Peter Mijnarends for important conversations. 
This work is supported in part by the US Department of Energy contract
W-31-109-ENG-38, and benefited from a travel grant from NATO and the 
allocation of supercomputer time at the NERSC and the Northeastern 
University Advanced Scientific Computation Center (NU-ASCC).

\end{multicols}

\begin{thebibliography}{99}

\bibitem{williams77} B. G. Williams,
		    {\it Compton Scattering} (McGraw-Hill, New York, 1977).
\bibitem{cooper85} M. J. Cooper,
		   Rep. Prog. Phys. {\bf 48}, 415(1985).
\bibitem{mijnarends95}  P. E. Mijnarends and A. Bansil
		    in {\it Positron Spectroscopy of Solids},
                    eds. A. Dupasquier and A. P. Mills,
                    Int. School of Physics 'Enrico Fermi',
                    pp. 257(1995).
\bibitem{bansil93} A. Bansil,
		   Z. Naturforsch. Teil A {\bf 48a},165 (1993).
\bibitem{bankap97} A. Bansil
                   and S. Kaprzyk, Mat. Science Forum {\bf 255-257}, 129(1997). 
\bibitem{sakurai95} Y. Sakurai, Y. Tanaka, A. Bansil, S. Kaprzyk,
		    A. T. Stewart, Y. Nagashima, T. Hyodo,
		    S. Nanao, H. Kawata, and N. Shiotani,
		    Phys. Rev. Lett. {\bf 74}, 2252 (1995).
\bibitem{schulke96} W. Sch\"ulke, G. Stutz, F. Wohlert, and A. Kaprolat,
		    Phys. Rev. {\bf B54}, 14381(1996).
\bibitem{jap1}      Y. Tanaka, A. Bansil, S. Kaprzyk, P. E. Mijnarends, 
                    Y. Sakurai, A. T. Stewart and N. Shiotani (preprint).
\bibitem{dob1}      L. Dobrzinski (preprint).
\bibitem{stutz99}  G. Stutz,  F. Wohlert, A. Kaprolat, W. Sch\"ulke,
                   Y. Sakurai, Y. Tanaka, M. Ito, H. Kawata, N. Shiotani,
                   S. Kaprzyk, and A. Bansil,
		   Phys. Rev. {\bf B60}, 7099(1999).
\bibitem{hamalainen96} K. H\"am\"al\"ainen, S. Manninen,
		       C.-C. Kao, W. Caliebe, J. B. Hastings,
		       A. Bansil, S. Kaprzyk, and 
		       P. M. Platzman,
		       Phys. Rev. {\bf B54}, 5453(1996).
\bibitem{itou97} M. Itou, Y. Sakurai, T. Ohata,
		 A. Bansil, S. Kaprzyk,
		 Y. Tanaka, H. Kawata, and N. Shiotani,
		 J. Phys. Chem.  Sol. {\bf 59}, 99(1998).
\bibitem{huotari00} S. Huotari, K. Hamalainen, S. Manninen, S. Kaprzyk, 
                    A. Bansil, W. Caliebe, T. Buslaps, V. Honkimaki, and 
                    P. Suortti (preprint)
\bibitem{suortti00} P. Suortti, T. Buslaps, V. Honkim\"aki, C. Metz, A. Shukla, 
    Th. Tschentscher, J. Kwiatkowska, F. Maniawski, A. Bansil, S. Kaprzyk, 
    A. S. Kheifets, D. R. Lun, T. Sattler, J. R. Schneider and F. Bell, 
    J. Phys. Chem. Solids {\bf 61}, 397(2000). 
\bibitem{japal}    T. Ohata, M. Itou, I. Matsumoto, S. Kaprzyk, 
                   P. E. Mijnarends, A. Bansil, Y. Sakurai, H. Kawata 
                   and N. Shiotani (preprint).
\bibitem{sakurai99} Y. Sakurai,  S. Kaprzyk,  A. Bansil, Y. Tanaka,
                    G. Stutz,
		    H. Kawata, and N. Shiotani,
   		    J. Phys. Chem.  Sol. {\bf 60}, 905(1999).
\bibitem{bansil98}  A. Bansil, S. Kaprzyk, A. Andrejczuk, L. Dobrzynski, 
                    J. Kwiatkowska, F. Maniawski, and E. Zukowski, 
                    Phys. Rev.  {\bf B57}, 314(1998).
\bibitem{mat00}   I. Matsumoto, J. Kwiatkowska, F. Maniawski, A. Bansil, 
     S. Kaprzyk, M. Itou, H. Kawata and N. Shiotani, 
     J. Phys. Chem. Solids {\bf 61}, 375(2000). 
\bibitem{kaprzyk90} S. Kaprzyk and A. Bansil,
		    Phys. Rev. {\bf B42}, 7358(1990).
\bibitem{bansil91}  A. Bansil and S. Kaprzyk,
		    Phys. Rev. {\bf B43}, 10335(1991).
\bibitem{bansil92}     A. Bansil, S. Kaprzyk, and J. Tobo{\l}a,
		       {\it Applications of Multiple Scattering
		        Theory in Material Science},
		        Mat. Res. Soc. Symp. Proc. {\bf 253}, 505(1992).
\bibitem{bansil99}  A. Bansil, S. Kaprzyk, J. Tobola and P. E. Mijnarends,
                    Phys. Rev. {\bf B60}, 13396(1999).
\bibitem{mijnarends76} P. E. Mijnarends and A. Bansil,
	               Phys. Rev. {\bf 13}, 2381 (1976).
\bibitem{bansil81}  A. Bansil, R.S. Rao, P. E. Mijnarends and L. Schwartz,
		    Phys. Rev. {\bf B23}, 3608 (1981).

\end{thebibliography}
\end{document}